\documentclass[a4paper,aps,prd,10pt,preprintnumbers,showpacs,twocolumn,superscriptaddress,nofootinbib,amsmath,amssymb]{revtex4-1}
\usepackage[dvips]{graphics}

\def\imo{i}
\def\re#1{Re(#1)}
\def\im#1{Im(#1)}

\begin{document}
\title{Detection of gravitational waves from black holes: Is there a window for alternative theories?}

\author{Roman Konoplya}
\email{konoplya@th.physik.uni-frankfurt.de}
\affiliation{Institute for Theoretical Physics, Goethe University,
  Max-von-Laue-Str. 1, 60438 Frankfurt, Germany}

\author{Alexander Zhidenko}
\email{zhidenko@th.physik.uni-frankfurt.de}
\affiliation{Institute for Theoretical Physics, Goethe University,
  Max-von-Laue-Str. 1, 60438 Frankfurt, Germany}
\affiliation{Centro de Matem\'atica, Computa\c{c}\~ao e Cogni\c{c}\~ao,
  Universidade Federal do ABC (UFABC), Rua Aboli\c{c}\~ao, CEP:
  09210-180, Santo Andr\'e, SP, Brazil}

\begin{abstract}

Recently the LIGO and VIRGO collaborations reported the observation of gravitational-wave signal corresponding to the inspiral and merger of two black holes, resulting into formation of the final black hole. It was shown that the observations are consistent with the Einstein theory of gravity with high accuracy, limited mainly by the statistical error. Angular momentum and mass of the final black hole were determined with rather large allowance of tens of percents. Here we shall show that this indeterminacy in the range of the black-hole parameters allows for some non-negligible deformations of the Kerr spacetime leading to the same frequencies of the black-hole ringing. This means that at the current precision of the experiment there remains some possibility for alternative theories of gravity.

\end{abstract}
\pacs{04.50.Kd,04.70.Bw,04.30.-w,04.80.Cc}

\maketitle


\section{Introduction}

A century after the formulation of General Relativity the LIGO and VIRGO collaborations \cite{Abbott:2016blz,TheLIGOScientific:2016src} detected gravitational waves from a pair of merging black holes and answer thereby a number of appealing questions related to our understanding of astrophysics, black holes, and gravitation. Interaction of two black holes can be conditionally divided into the four stages:
\begin{enumerate}
\item Newtonian stage, when the distance between two black hole is much larger than their sizes; it includes rotation of the black holes around each other in close orbit, \emph{inspiral} \cite{Buonanno:2006ui,Berti:2007fi};
\item the merger of two black holes into a single one which ends up with;
\item the ringdown phase characterized by the \emph{quasinormal modes} \cite{reviewsQNMs} of the resultant black hole.
\end{enumerate}
The last stages of formation of a single black hole and the consequent quasinormal ringing, corresponding to the regime of a strong gravitational field, cannot be described in terms of the post-Newtonian approximation. These last stages represent essential intrinsic characteristics of a theory of gravity.

Indeed, there is a number of alternative theories of gravity which produce the same black-hole behavior at a far distance from its surface, i.e. in the asymptotic region, but lead to qualitatively different features near the event horizon. One of the aims of detection of gravitational waves from black holes is testing the black-hole near-horizon geometry and distinguishing Kerr spacetime \cite{Kerr:1963ud} from, possibly, another black-hole geometry corresponding to some alternative theory of gravity\footnote{In some cases the Kerr metric can also be a solution of non-Einsteinian theories of gravity, for example, of some $f(R)$ theories, though the perturbations equations and, thereby, the ringdown profile will be different from the Einsteinian one \cite{Psaltis:2007cw,Barausse:2008xv}.}.

Comparison of numerical simulations of the gravitational-wave signal, made within the Einstein gravity, with the observations (fig. 1 in \cite{Abbott:2016blz}) shows very good agreement up to a few percents.  However, there is a rather large range of possible values of mass and angular momentum of the black hole (see fig. 3 in \cite{TheLIGOScientific:2016src}) leading to the same gravitational-wave signal within the achieved accuracy. \emph{This range of allowed values of the black-hole parameters could  naturally be imagined as opportunity for deviation from the Kerr spacetime instead of deviation from given values of black-hole parameters within the same Kerr geometry.} This intuitive thought is supported by understanding that the quasinormal frequencies strongly depend on mass and angular momentum of a black hole, so that two black holes with different masses and momenta in two different theories of gravity may produce very close dominant quasinormal frequencies. If it is so, agreement of the observed gravitational wave signal with the one predicted by General Relativity (GR) for the Kerr spacetime would rule out all alternatives only if parameters of the final black hole were determined with high accuracy (and, preferably, independently on the supposition of the validity of GR) and shown to be equal to the Kerr's ones. At the moment this is not the case, though the precision of the experiment will be increasing in the near future, what should, one way or the other, give us more constrained range of the black-hole parameters.

Here we shall show that the current indeterminacy in the values of the black-hole parameters allows for non-negligible deviations from the Kerr spacetime which leads to essentially the same quasinormal ringing. This may mean that not only the Einsteinian theory of gravity is consistent with the latest observations of gravitational waves, but also some deviations from it do not contradict the ringdown picture.

For this purpose in Sec. II we shall ``prepare'' a rather arbitrary deformation of the Kerr spacetime, which preserves asymptotic properties of the Kerr metric, such as its post-Newtonian expansion coefficients, relation between quadrupole momentum and mass, but drastically changes its near-horizon behavior. For simplicity, the deformation is described by only one parameter, which is also fully justified by purely illustrative aim of our note. Then, we shall show that the large indeterminacy in $a$ and $M$ of such a deformed black hole allows for a wide range of values of the deformation parameter. In Sec. III we shall give another example: the Kerr black hole with a fixed angular momentum will be shown to produce quasinormal modes which are close to those of the Sen black hole with different value of the angular momentum and a nonzero dilaton.

\section{Kerr vs deformed Kerr space-time}

For convenience, we shall consider the line element of an arbitrary axially symmetric black hole in the following form \cite{Konoplya:2016jvv}
\begin{eqnarray}\label{metric}
ds^2 &=&
-\dfrac{N^2(r,\theta)-W^2(r,\theta)\sin^2\theta}{K^2(r,\theta)}dt^2\\
&&-2W(r,\theta)r\sin^2\theta dtd\phi+K^2(r,\theta)r^2\sin^2\theta d\phi^2\nonumber\\
&&+\Sigma(r,\theta)\left(\dfrac{B^2(r,\theta)}{N^2(r,\theta)}dr^2 + r^2d\theta^2\right),\nonumber
\end{eqnarray}
where the Kerr metric is given as
\begin{eqnarray}\nonumber
N^2(r,\theta)& = &\frac{r^2-2Mr+a^2}{r^2}\,,\\
\nonumber
B(r,\theta)& = &1\,,\\
\label{Kerrfunc}
\Sigma(r,\theta)& = &\frac{r^2+a^2\cos^2\theta}{r^2}\,,\\
\nonumber
K^2(r,\theta)& =
&\frac{(r^2+a^2)^2-a^2\sin^2\theta
(r^2-2Mr+a^2)}{r^2(r^2+a^2\cos^2\theta)}\,,\\
\nonumber
W(r,\theta)& = &\frac{2Ma}{r^2+a^2\cos^2\theta}\,,
\end{eqnarray}
where $M$ is the mass and $a$ is the rotation parameter.

Now, we shall deform the above Kerr spacetime by adding a static deformation which changes the relation between the black-hole mass and position of the event horizon, but preserves asymptotic properties of the Kerr spacetime. Namely, the substitution
\begin{equation}
M\rightarrow M+\dfrac{\eta}{2r^2},
\end{equation}
once it is used in (\ref{Kerrfunc}), modifies the Kerr metric as follows

\begin{eqnarray}\nonumber
N^2(r,\theta)& = &\frac{r^2-2Mr+a^2}{r^2}-\frac{\eta}{r^3}\,,\\
\nonumber
B(r,\theta)& = &1\,,\\
\label{Modiffunc}
\Sigma(r,\theta)& = &\frac{r^2+a^2\cos^2\theta}{r^2}\,,\\
\nonumber
K^2(r,\theta)& =
&\frac{(r^2+a^2)^2-a^2\sin^2\theta
(r^2-2Mr+a^2)}{r^2(r^2+a^2\cos^2\theta)}\\
\nonumber
&&+\frac{a^2\eta\sin^2\theta}{r^3(r^2+a^2\cos^2\theta)}\\
\nonumber
W(r,\theta)& = &\frac{2Ma}{r^2+a^2\cos^2\theta}+\frac{\eta a}{r^2(r^2+a^2\cos^2\theta)}\,,
\end{eqnarray}
where $M$ is the ADM mass and $a=J/M$ is the rotation parameter.

The above constructed spacetime of the deformed black hole possesses the following important for us properties:
\begin{enumerate}
\item it allows for the separation of radial and angular variables in the field equation, what allows us to reduce the perturbation problem to a radial, master wave-like equation,
\item it has the same post-Newtonian asymptotic ($\beta=\gamma=1)$ as the Kerr metric,
\item the quadrupole momentum of such a deformed spacetime obeys the same relation $Q=-Ma^2$ as the Kerr black hole,
\item the deformed metric has quite different (from Kerr) near-horizon geometry, expressed, in particular, in a different position of the spherical event horizon.
\end{enumerate}

Similarly to the Kerr black hole, the Killing horizon obeys the following relation,
\begin{equation}
g^{rr}\equiv \frac{N^2(r,\theta)}{B^2(r,\theta)}=\frac{r^2-2Mr+a^2}{r^2}-\frac{\eta}{r^3}=0,
\end{equation}
and coincides with the event horizon.

It is convenient to parametrize the considered family of metrics with an additional parameter $r_0$, so that
$$\eta=r_0(r_0^2-2Mr_0+a^2).$$
We shall use the parameter $\delta r$ measuring deviation of the position of the event horizon from the Kerr one $r_\mathrm{Kerr}$,
$$r_0=r_\mathrm{Kerr}+\delta r=M+\sqrt{M^2-a^2}+\delta r,$$
so that $\delta r=0$ implies $\eta=0$ and corresponds to the Kerr metric. Although $\delta r$ is a coordinate dependent measure of deviation from the Kerr geometry, the coordinates (\ref{Modiffunc}) are indistinguishable from the Boyer-Lindquist coordinates at infinity, so that different distant stationary observers should in principle agree on what is accepted as ``large deviation from Kerr''.

As our aim is only to \emph{evaluate the order of an allowed range of the deformation parameter} $\delta r$ at a given relatively small allowance for the  quasinormal frequency (a few percents), we do not need to be tied to a particular theory, type of perturbation or even fixed value of the quasinormal frequency. Therefore, we shall consider a test scalar field in the deformed background (\ref{Kerrfunc}) and  use simple semi-classical WKB estimates. Such a test-field approach will not distinguish the Kerr space-time as a solution of the Einstein field equations from the same Kerr space-time as solution of some non-Einsteinian gravity mentioned above \cite{Psaltis:2007cw,Barausse:2008xv}. However, our aim here is not to include all the possible alternative theories into consideration, but to show that at least \emph{some deviations} from the Einstein gravity are still allowed by the observations. Analysis of gravitational perturbations would obviously constrain the possible alternatives better.

A massless minimally-coupled scalar field obeys the equation
\begin{equation}\label{boxphi}
\Phi^{;\mu}_{;\mu}=\frac{1}{\sqrt{-g}}\partial_{\mu}\left(\sqrt{-g}g^{\mu\nu}\partial_{\nu}\Phi\right)=0.
\end{equation}

Substituting the ansatz
$$\Phi(t,r,\theta,\phi)=\exp(-\imo\omega t+\imo m \phi)R(r)S(\theta)(r^2+a^2)^{-1/2}\,,$$
into (\ref{boxphi}) we find that $S(\theta)$ satisfies the equation for the spheroidal functions
\begin{eqnarray}\nonumber
\left(\frac{d^2}{d \theta^2} + \cot \theta \frac{d}{d\theta} - \frac{m^2}{\sin^2 \theta} - a^2 \omega^2\sin^2 \theta + \lambda_{\ell m}(\omega) \right) S(\theta)\\ = 0\,,\quad \label{angularpart}
\end{eqnarray}
where the values of the separation constant $\lambda_{\ell m}(\omega)$ can be enumerated, for each given integer azimuthal number $m$, by the multipole number
$$\ell=|m|,|m|+1,|m|+2,\ldots\,.$$

For the radial function $R(r)$ we obtain the wave-like equation
$$\frac{d^2R}{dr_\star^2}+\left(\omega^2-V(r,\omega)\right)R(r)=0,$$
where
$$dr_\star=\left(\frac{r^2-2Mr+a^2}{r^2+a^2}-\frac{\eta}{r^3+ra^2}\right)^{-1}dr,$$
is the tortoise coordinate, and the effective potential is given by the following relation
\begin{eqnarray}
V(r,\omega)&=&\frac{2 a m (2 M r^2 + \eta)}{r(r^2+a^2)^2}\omega-\frac{m^2a^2}{(r^2+a^2)^2}\\\nonumber
&&+\left(\frac{r^2-2Mr+a^2}{r^2+a^2}-\frac{\eta}{r^3+ra^2}\right)\times
\\\nonumber
&&\hspace{-2em}\Biggr(\frac{\lambda_{\ell m}(\omega)}{r^2+a^2}+\frac{a^4+a^2 r^2-4Ma^2r+2 M r^3+3 r \eta}{(r^2+a^2)^3}\Biggr)\,.
\end{eqnarray}

\begin{figure}
\resizebox{\linewidth}{!}{\includegraphics{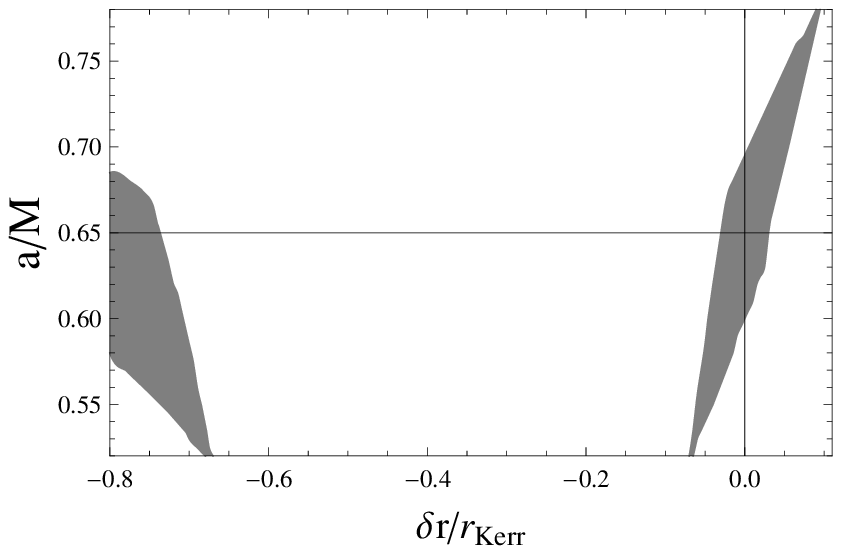}}
\caption{Parametric region (gray) of possible deformations $\delta r/r_\mathrm{Kerr}$ leading to the ringdown frequency $\omega M=0.635-0.0901\imo$ (which corresponds, according to the WKB formula for the Kerr metric with $a/M\approx0.65$) within $3\%$ accuracy.}\label{fig:errorreg}
\end{figure}

Application of the WKB formula \cite{WKB}  at fixed values of $\omega$
\begin{equation}
\omega^2 = Q (\ell, m, \omega, M, a, \delta r),
\end{equation}
where the explicit form of the operator $Q$ depends on the order of the WKB series, allows one to find those values of the deformation parameter $\delta r$, which, in the allowed indeterminacy range for $a$ and $M$, reproduce the quasinormal frequency $\omega$ within the desired few percents accuracy. We have chosen $\ell=m=2$ mode for Kerr and the rate of rotation about $a/M \approx 0.65$ and tested possible values of $\delta r$ (see fig.~\ref{fig:errorreg}) against the allowed range of $a/M$ determined in fig.~3 of \cite{TheLIGOScientific:2016src}. From fig.~\ref{fig:errorreg}  we can see that \emph{the deformation $\delta r/r_{Kerr}$ can achieve several tens of percents}. Although the particular shape of the region depicted in fig.~\ref{fig:errorreg} depends on the spin of perturbation, type of the chosen deformation and a number of computational details of quasinormal modes, the above statement on the \emph{order} of allowed deformation from Kerr geometry evidently must not depend on any of these details. Indeed, a reader could repeat our computations for vector and spinor fields, as well as choose a different value of $\omega$ for comparison. The analysis of dependence of quasinormal modes for a great variety of black holes \cite{reviewsQNMs} shows that the order of ``sensitivity'' of $\omega$ as to the change of the black hole parameters is the same for gravitational perturbations as for perturbation of test fields.

In the next section we shall give another illustration of the same idea and go over from the ``ad-hoc deformation'' to consideration of the particular alternative theory, Einstein-dilaton gravity, where the nonzero dilaton parameter $b$ plays the role of deformation.

\section{Kerr-Sen vs Kerr black holes}

It is natural to expect that determination of not only a single mode (that could ``by accident'' be close to the corresponding mode in some alternative theory), but one or more subdominant frequencies would help to identify the black hole geometry much easier \cite{Bertietc}. Here, we shall consider another example: a comparison between scalar quasinormal frequencies of Kerr and Kerr-Sen black holes. We shall show that, quite counter-intuitively, knowing of a few higher modes within a few percents accuracy does not remedy the situation, leaving the above discussed indeterminacy.

The Sen black hole is a rotating, charged black hole in the four-dimensional heterotic string theory which can be described by the line element
\begin{eqnarray}
ds^2 &=& \frac{\Delta_r}{\Sigma}(dt-a\sin^2\theta d\varphi)^2-\Sigma
\left(\frac{dr^2}{\Delta_r}+d\theta^2\right)\\\nonumber
&&-\frac{\Delta_\theta \sin^2\theta}{\Sigma}
[adt-(r^2+2br+a^2)d\varphi]^2,
\end{eqnarray}
where
\begin{eqnarray}\nonumber
\Delta_r&=&r^2-2(M-b)r+a^2,\\
\Sigma&=&r^2+2br+a^2\cos^2\theta,\nonumber
\end{eqnarray}
$a$ is the rotation parameter, $M$ is the ADM mass and $M>b\geq0$. The Maxwell and dilaton fields are given by
\begin{eqnarray}
A_{\mu}dx^{\mu}&=&Q\frac{r}{\Sigma}(dt-a\sin^2\theta d\varphi),\\
e^{2\phi}&=&W\frac{r^2+a^2\cos^2\theta}{\Sigma},
\end{eqnarray}
where the electric charge $Q$ is related to $b$ as
\begin{equation}
Q^2=2WMb.
\end{equation}

\begin{table}
\begin{tabular}{|l|l|l|l|l|}
\hline
$n$&$a=0.65M$, $b=0$&$a=0.55M$, $b=0.1M$&$\re\%$&$\im\%$\\
\hline
$0$&$0.635603\!-\!0.0896663\imo\!$&$0.637270\!-\!0.0914843\imo\!$&$0.262\%\!$&$2.028\%\!$\\
$1$&$0.625909\!-\!0.2709374\imo\!$&$0.627168\!-\!0.2765441\imo\!$&$0.201\%\!$&$2.069\%\!$\\
$2$&$0.608566\!-\!0.4573669\imo\!$&$0.609178\!-\!0.4671567\imo\!$&$0.101\%\!$&$2.140\%\!$\\
$3$&$0.586811\!-\!0.6504648\imo\!$&$0.586793\!-\!0.6649052\imo\!$&$0.003\%\!$&$2.220\%\!$\\
$4$&$0.563709\!-\!0.8498708\imo\!$&$0.563277\!-\!0.8693261\imo\!$&$0.077\%\!$&$2.289\%\!$\\
\hline
\end{tabular}
\caption{Quasinormal modes for Kerr ($b=0$) and Kerr-Sen ($b\neq0$) black holes $\omega M$ for $\ell=m=2$. Last two columns give the difference between Kerr and Kerr-Sen quasinormal frequencies in percents.}\label{tabl:l=2}
\end{table}

\begin{table}
\begin{tabular}{|l|l|l|l|l|}
\hline
$n$&$a=0.65M$, $b=0$&$a=0.55M$, $b=0.1M$&$\re\%$&$\im\%$\\
\hline
$0$&$0.904059\!-\!0.0893540\imo\!$&$0.904496\!-\!0.0911598\imo\!$&$0.048\%\!$&$2.021\%\!$\\
$1$&$0.897288\!-\!0.2690321\imo\!$&$0.897398\!-\!0.2745294\imo\!$&$0.012\%\!$&$2.043\%\!$\\
$2$&$0.884493\!-\!0.4514534\imo\!$&$0.884014\!-\!0.4608656\imo\!$&$0.054\%\!$&$2.085\%\!$\\
$3$&$0.867011\!-\!0.6379172\imo\!$&$0.865805\!-\!0.6515619\imo\!$&$0.139\%\!$&$2.139\%\!$\\
$4$&$0.846474\!-\!0.8290498\imo\!$&$0.844545\!-\!0.8472661\imo\!$&$0.228\%\!$&$2.197\%\!$\\
\hline
\end{tabular}
\caption{Quasinormal modes for Kerr ($b=0$) and Kerr-Sen ($b\neq0$) black holes $\omega M$ for $\ell=m=3$. Last two columns give the difference between Kerr and Kerr-Sen quasinormal frequencies in percents.}\label{tabl:l=3}
\end{table}

Quasinormal modes were computed with the help of the Leaver method \cite{Leaver:1985ax} in \cite{Konoplya:2006br} for Kerr and in \cite{Kokkotas:2015uma} for Kerr-Sen black holes. From table~\ref{tabl:l=2} we can see that the $\ell=m=2$ quasinormal modes for Kerr black hole with $a=0.65M$ and for Sen black hole with $a=0.55M$ and the value of the dilaton parameter $b=0.1M$ are as close as one percent at the real oscillation frequency and as about two percents at the damping rate. This occurs not only for the fundamental mode, but also for higher overtones. In addition, it takes place for higher multipoles $\ell$ as can be seen from table~\ref{tabl:l=3}. Therefore, if in the future we see the coalescence of two black holes of considerably different masses, so that $\ell=3, 4$ modes are highly excited, that probably would not remedy the situation unless the precision of the determination of the black hole's parameters will be greatly improved. Notice, that the Kerr-Sen metric is considered here only as another toy model for an illustration of the parametric indeterminacy within an alternative theory, because the considered value of the parameter $b$ is not realistic and implies a much larger electric charge than an astrophysical black hole can possess.

\section{Conclusion}

Using simple semi-classical arguments as well as numerical data for quasinormal modes of various black holes, in this letter we have shown that the indeterminacy with which gravitational-wave signal constrains the mass and angular momentum of the black hole allows not only the Kerr spacetime to be consistent with the gravitational ringdown profile, but also leaves a window for non-negligibly deformed (from Kerr) spacetimes with the same asymptotic properties. As mass and angular momentum of the system are measured by comparison with simulations of the earlier stages of the black holes' interaction, that is before the final ringdown, these parameters are found within the post-Newtonian formalism at some order. Thus, one could admit that there might exist a strongly deformed Kerr-like black hole, corresponding to an alternative theory of gravity, such that its behavior in the post-Newtonian regime is quite similar to Kerr black hole, while its near-horizon behavior is different. Here we have illustrated this idea with the help of WKB computations, as shown in fig.~\ref{fig:errorreg}. Another, even simpler illustration of the same idea has been given here through comparison of the quasinormal spectra of Kerr and Kerr-Sen black holes at different values of black hole parameters. It has been shown that the Kerr black hole with $a= 0.65 M$ produces quasinormal modes which are very close to the ones of rotating dilaton black holes with $a= 0.55 M$ and not negligible value of the dilaton parameter $b=0.1M$. All the values of $a/M$ are taken in the range allowed by the observation of gravitational-waves \cite{Abbott:2016blz}, while the difference in the quasinormal spectra of Kerr and dilatonic black holes is less than the accuracy of determination of the detected quasinormal mode. Quite unexpectedly, this ``proximity'' of quasinormal spectra occurs not only for a single mode $\ell=m=2$, but also for higher modes as well as for other multipoles $\ell$.

The comparison of the detected gravitational wave profile at the inspiral and merger phases which requires post-Newtonian approach at high orders will definitely constrain our freedom until some extend. Though constraints given at these earlier stages (see table~1 of \cite{TheLIGOScientific:2016src}) are quite loose and, apparently, should not change our conclusions qualitatively.

In order to disprove the above proposal, one needs to determine the black-hole parameters with high accuracy. In the future this could be done either by improving the accuracy of detection of the gravitational-wave profile or with complementary observations of black holes in the electromagnetic spectrum, which could potentially give us an image of a black hole.

This letter in no way contradicts the Einstein theory of gravity or cast shadows on the great importance of the recent outstanding discovery of gravitational waves. In essence, we simply show that the current experimental allowance in the determination of the black hole parameters can be interpreted as freedom for alternative theories of gravity as well.

\begin{acknowledgments}
This work was partially supported by Conselho Nacional de Desenvolvimento Cient\'ifico e Tecnol\'ogico (CNPq).
\end{acknowledgments}

\end{document}